\documentclass{sig-alternate-10pt}
\usepackage{url}
\usepackage[backgroundcolor=yellow]{todonotes}
\presetkeys{todonotes}{inline}{}

\usepackage[
  pass,
]{geometry}

\usepackage{pgfplots}
\pgfplotsset{compat=1.9}

\tikzset{
      font={\fontsize{9pt}{10}\selectfont}}

\usepackage[font=footnotesize,labelfont=bf]{caption}

\begin{document}

\title{TCPSnitch: Dissecting the Usage of the Socket API}


\newcommand{\tcpsnitch}{\texttt{TCPSnitch} }
\newcommand{\tcpsnitchns}{\texttt{TCPSnitch}}

\numberofauthors{1}
\author{
\alignauthor Gregory Vander Schueren\textsuperscript{1}, Quentin De Coninck\textsuperscript{2} and Olivier Bonaventure\textsuperscript{2} \\
    \affaddr{Universite Catholique de Louvain, Louvain-la-Neuve, Belgium}\\
    \email{\textsuperscript{1}gregory.vanderschueren@gmail.be, \textsuperscript{2}\{first.last\}@uclouvain.be}
}
\maketitle

\begin{abstract}
Networked applications interact with the TCP/IP stack through
the socket API. Over the years, various extensions
have been added to this popular API. In this paper, we
propose and implement the \tcpsnitch software that
tracks the interactions between Linux and Android applications
and the TCP/IP stack. We collect a dataset containing the
interactions produced by more than 120 different applications.
Our analysis reveals that applications use a variety of
API calls. On Android, many applications use various socket
options even if the Java API does not expose them directly.
\tcpsnitch and the associated dataset are publicly available.
\end{abstract}

\section{Introduction}

The socket API was introduced together with release 4.2 of
the BSD Unix distribution that included a functional TCP/IP stack
\cite{Quarterman:BSD}. This API allows applications to interact with
the underlying networking stack. When the socket API
was designed, TCP/IP was one family of network protocols among many
others and it was important to abstract those protocol families.
The heart of the socket API is  a set of basic system calls
including \texttt{socket}, \texttt{bind}, \texttt{connect},
\texttt{accept}, \texttt{close}, \texttt{listen}, \texttt{send},
\texttt{receive},\ldots Those system calls interact with the
underlying network implementation that is part of the operating
system kernel. The socket API was not the only approach to
interact with the network stack. The STREAM API, based on
\cite{ritchie1984unix} was extended to support the TCP/IP protocol
stack and used in Unix System V \cite{rago1993unix}.

Over the years, the popularity of the socket API grew in parallel
with the deployment of the global Internet. Nowadays, the
socket API is considered by many as the standard API for
simple networked applications. Several popular textbooks are entirely
devoted to this API \cite{Stevens:UNP,donahoo2001pocket}. Given the
importance of web-based applications, many developers do not interact
directly with the socket API anymore but rely on higher-level
abstractions. For example, programming languages such as Java or Python
include libraries exposing URL and implementations of
HTTP/HTTPS. For C developers, libraries such as \texttt{libcurl} 
also provide higher level abstractions.

During the last 30 years, the socket API has evolved, with new features added
over the years. Some were dedicated to the support of specific features, such
as ATM \cite{almesberger1998using} or Quality of Service
\cite{abbasi2002quality}. Other extensions \cite{Provos_Scalable:2000,
gammo2004comparing}
focused on improving the
interactions between applications and the underlying stack through
\texttt{poll/select}, \texttt{epoll},\ldots On Linux, new system
calls to directly send pages or entire files (like \texttt{sendfile})
were added. Furthermore, socket extensions have been defined
for each new transport protocol \cite{schier2012using,natarajan2009sctp}.
Socket extensions have also been proposed to deal with multihoming
\cite{Schmidt:intents} and specific APIs have been implemented on top
of Multipath TCP \cite{hesmans2015smapp,hesmans2016enhanced}.

Recently, the Internet Engineering Task Force created the 
Transport Services (taps) working group whose main objective is
\emph{to help application and network 
stack programmers by describing an (abstract) interface for applications 
to make use of Transport Services}. Although this work will focus on
abstract transport services, understanding how the current APIs
are used by existing applications will help in designing generic transport
services that correspond to their needs.

We propose \texttt{TCPSnitch}, an open-source software that collects
detailed traces of the interactions between networked applications and
the Linux TCP/IP stack and sends them to a publicly available database exposing
various statistics. This paper is organized as follows. We first
describe \tcpsnitch in section~\ref{sec:tcpsnitch}. Then we present in
section~\ref{sec:dataset} the traces that we collected from 90
different Android applications. Section~\ref{sec:udp} analyses in more
details the utilization of UDP sockets by those applications while
section~\ref{sec:tcp} focuses on the TCP sockets. We summarize the
main findings of this work and future work in section~\ref{sec:conclusion}.

\section{TCPSnitch}\label{sec:tcpsnitch}

Different solutions have been proposed and implemented to analyze the
utilization of system and library calls by applications. Two
approaches are possible. The first one is to analyze the
application code (binary or sometimes source for open-source
applications) and extract the interesting calls from the corresponding
files. Several researchers have adopted this approach to study
networked applications. In 2011,~\cite{Komu:Sockets} analyzed the source
code of 2187 Ubuntu applications to detect the presence of certain keywords
of the socket API. In 2016,~\cite{Tsai:LinuxAPI} disassembled
binaries of 30K Linux applications using \texttt{objdump} and performed
a call-graph analysis to study the Linux API usage. Still in 2016,
~\cite{Atlidakis:POSIX} proposed \texttt{libtrack} and analyzed 1.1M Android
applications for linkage with POSIX functions. The main advantage of this
approach is that it is possible to analyze a large number of applications to
determine the system calls used by the majority of the applications.
Unfortunately, it is very difficult to determine
which parameters are passed to these identified functions or how
frequently they are called. Source code analysis is also impractical for
closed-source applications.

The second approach is to instrument the application and intercept the
system or library calls. On Unix variants, the \texttt{strace} or
\texttt{ltrace} applications can be used to collect traces of the
system or library calls. The \texttt{libtrack} tool proposed by
\cite{Atlidakis:POSIX} also supports dynamic tracing of functions invocations.
\tcpsnitch currently intercepts 40 functions that are related to the network
stack. \tcpsnitch tracks the functions that are applied on each
socket with their timestamp, parameters and return value.
It also collects metadata information such as system information,
the network configuration and the kernel version. Compared to simpler
tools like \texttt{strace} and \texttt{libtrack}, a major benefit of
\tcpsnitch is that all the data collected during the utilization of an
application can be uploaded on a public database. The web interface of this 
database, available on \texttt{https://tcpsnitch.org}, provides different
visualizations of the database and allow users to browse through the
collected traces. \tcpsnitch is written in C and
counts about 6500 lines of code, without blanks and comments.

Compared to the first approach, the main advantage of \tcpsnitch is
that it can trace sequences of calls and also collect information
about the function parameters and the return values. This enables us
to observe how and when socket API calls and options are used, the size of the
buffers used by \texttt{send()/recv()}, which thread called the API function,\ldots
While static analysis tools such as~\cite{Komu:Sockets} or~\cite{Tsai:LinuxAPI}
give indications about the possible usage of some socket API calls or options,
\tcpsnitch allows observing their actual use.

Since \tcpsnitch uses \texttt{LD\_PRELOAD} like \texttt{strace} to intercept
the functions calling the system calls in the standard
C library, it is possible for applications to bypass \tcpsnitch by
either being statically linked with the C library or directly using
the system calls. Before analyzing the results, it is important 
to note one caveat about the utilization of
\tcpsnitch on Android applications. On Android,
applications do not usually call \texttt{exit()} because they typically
remain running or idle once started. To end the tracing of an Android application,
\tcpsnitch calls the \texttt{force-stop} command of the activity
manager tool (\texttt{am}) to terminate the application. This means
that the application does not get the opportunity to cleanly close its
opened sockets. This caveat only affects the interception of the
\texttt{close()} function, not other functions.

To preserve the user privacy, he/she can opt-out for the collection of sensitive
metadata. \tcpsnitch does not trace the utilization of the DNS libraries and
thus does not collect domain names. With \texttt{send()/recv()},
\tcpsnitch only collects the buffer pointers and sizes, not the actual
data. Furthermore, all the non-loopback or link-local IP addresses that are
collected as parameters of the traced system calls are replaced by the low order
32 (resp. 128) bits of a SHA-1 hash computed over the concatenation of a
random number generated by \tcpsnitch when it starts and the IP address.

\section{Dataset}\label{sec:dataset}

Using \tcpsnitchns, we recorded traces for 90 Android and 33 Linux
applications by manually interacting with each application for a few seconds
in order to reproduce a typical usage. We mainly selected popular
consumer-oriented client applications and the dataset currently does not
include any server-side application. For some popular applications, we recorded
multiple traces in different network environments.

We observed major differences in the API usage patterns on Android and Linux.
For instance, the most popular API functions differ and applications use
different recurring combinations of socket options. In accordance with
\cite{Atlidakis:POSIX}, we confirm that high-level frameworks and libraries
drive the API usage and are at the root of such glaring
disparities. Due to space limitations, we restrict our analysis to the
Android dataset for the rest of this paper. The full dataset with
various visualizations is publicly available from \url{https://tcpsnitch.org}.

Our Android dataset mostly includes highly popular applications from different
categories of apps in the Google Play Store. Table~\ref{tab:apps} shows a
sample of representative applications. At the time of writing, all Android
traces have been recorded on Android 6.0.1 with a LG Nexus 5 device.
In total, the Android dataset includes 181 application traces
that opened a total of 16.384 sockets. This represents about 2.3M intercepted
function calls.

\begin{table}[]
    \centering
    \begin{tabular}{ll}
        \hline
        \multicolumn{1}{|l|}{\textbf{Category}} & \multicolumn{1}{l|}{\textbf{Applications}} \\ \hline
        Social                                  & Facebook, Twitter, Linkedin                \\
        Streaming                               & Spotify, Netflix, Soundcloud               \\
        Video-telephony                         & Skype, Viber, Hangout                      \\
        Shopping                                & Amazon, AliExpress,  Zalando               \\
        Browsers                                & Chrome, Firefox, Opera                     \\
        Productivity                            & Evernote, Slack, Mega                      \\
        Video/photo                             & Youtube, Instagram, Pinterest
    \end{tabular}
    \caption{Sample applications. \textmd{The Android dataset contains traces
of 90 applications from different categories of apps in the Google Play Store.}}
    \label{tab:apps}
\end{table}

\subsection{Usage of the socket API functions}

\begin{figure}
\centering
\includegraphics[width=\columnwidth]{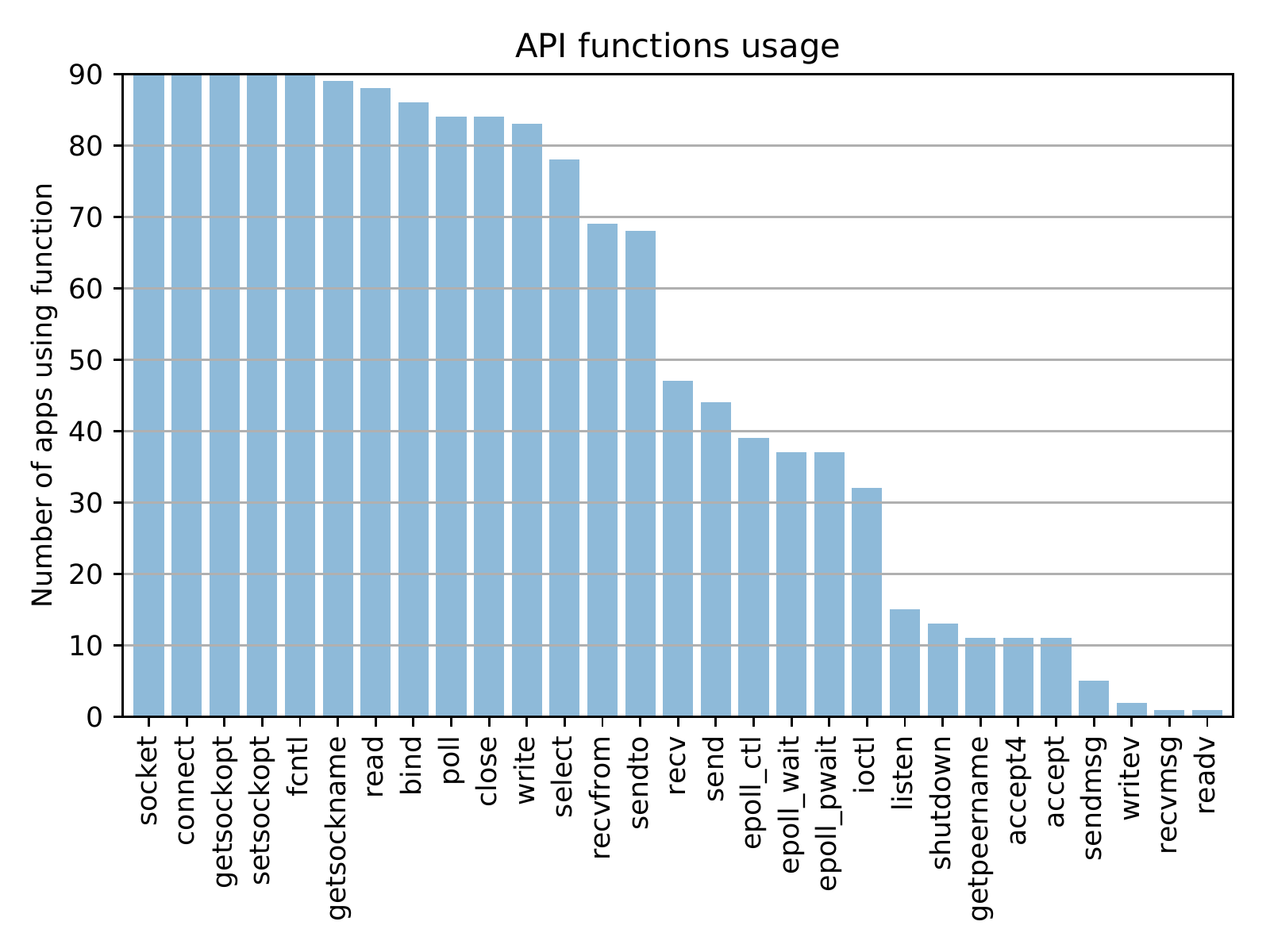}
\caption{API functions usage. \textmd{A dozen API functions are used by almost
all applications. Vectored I/O functions are mostly unused.}}
\label{fig:functions_usage}
\end{figure}

The socket API contains various functions that often have overlapping
purposes. For instance, there are as many as 7 functions to send data:
\texttt{write()}, \texttt{send()}, \texttt{sendto()}, \texttt{writev()},
\texttt{sendmsg()},\texttt{sendmmsg()} and\\\texttt{sendfile()}.
Figure~\ref{fig:functions_usage} shows the number of applications using
each intercepted function. This section intends to
shed some light about the real usage of these functions.

Some functions are used by a large fraction of the applications. For instance,
\texttt{getsockopt()}, \texttt{setsockopt()} and \texttt{fnctl()}
are used by all the applications in our dataset and only one
application does not call \texttt{get\-sock\-name()}. Another surprising result
is that a textbook server-side function such as
\texttt{bind()} is used by 96\% of our client Android
applications.  We observe that about 95\% of
these \texttt{bind()} calls specify \texttt{INADDR\_ANY} as the IP address and
0 for the port number (meaning an OS assigned random port) but explicitly request
for an IPv6 address. This usage is mainly driven by the \texttt{Socket} class
of the Android SDK~\cite{aosp_socket} that caches the local address of the
socket (using \texttt{get\-sock\-name()}) before trying to connect it.

Some of our observations are dependent on Android 6.0.1. For instance,
Bionic, Google's implementation of \texttt{libc},
implements some API functions by calling their more complex sibling,
e.g. \texttt{send()} is implemented by calling \texttt{sendto()}. When the
simple version of these 'twin' functions \footnote{Here is a complete list of
these twin functions: \texttt{send()} calls \texttt{sendto()}, \texttt{recv()}
calls \texttt{recvfrom()}, \texttt{accept()} calls \texttt{accept4()} and
\texttt{epoll\_wait} calls \texttt{epoll\_pwait()}.} is called, \tcpsnitch
records 2 consecutive function calls although the application code actually
performs a single function call. This means that the popularity of
\texttt{sendto()}, \texttt{recvfrom()}, \texttt{accept4()} and
\texttt{epoll\_pwait()} is overestimated on Fig.~\ref{fig:functions_usage}.

We did not observe any utilization of \texttt{sendfile()}, \texttt{sendmmsg()}
and \texttt{recvmmsg()}. These 3 functions are optimizations
mostly useful for server-side applications requiring high-performance.
For instance, \texttt{sendfile()} is a Linux specific call that saves 
a back-and-forth copy between kernel
and user space when sending a file over a socket, while \texttt{sendmmsg()} and
\texttt{recvmmsg()} allow to send or receive multiple \texttt{struct msghdr}
in a single system call.

Figure~\ref{fig:functions_usage} also shows that vectored I/O functions such as
\texttt{readv()} and \texttt{recvmsg()} are seldom used an Android.

\subsection{Types of sockets}

In the IPv6 enabled WiFi network used for the experiments, all but one application
established a TCP connection with a remote host over IPv6. This is a
confirmation of the growing importance of IPv6. All the surveyed
applications opened at least one IPv6 socket while only 64\% opened an IPv4
socket.

While all applications use asynchronous sockets, a single application used the
\texttt{SOCK\_NONBLOCK} optional flag when calling \texttt{socket()}.
\texttt{SOCK\_CLOEXEC} was never used. Most sockets are made asynchronous after
their creation using \texttt{fcntl(F\_SETFL)} and 5 applications used
\texttt{ioctl(FIONBIO)}. Usually, TCP sockets are turned asynchronous just
before the \texttt{connect()} call. As a matter a fact, \texttt{O\_NONBLOCK} is
the only file status flag used by the studied Android applications with
\texttt{fnctl(F\_SETFL)} and \texttt{fnctl(F\_GETFL)}. The \texttt{O\_APPEND},
\texttt{O\_ASYNC}, \texttt{O\_DIRECT} and \texttt{O\_NOATIME} flags were never
used.

\section{UDP sockets}\label{sec:udp}

We first analyze how UDP is used by the 31 applications in our dataset that
open at least one \texttt{SOCK\_DGRAM} socket. Note that those UDP sockets are
explicitly requested by the applications themselves since \tcpsnitch does not
track \texttt{getaddrinfo()} and related functions that are part of
\texttt{libc}.

\begin{table}[]
\centering
\begin{tabular}{ll}
85\% & Get info about network with \texttt{ioctl()} \\
8\% & \texttt{connect()} but does not exchange data \\
6\% & Send or receive data \\
1\% & Other usages
\end{tabular}
\caption{UDP sockets usage. \textmd{Most UDP sockets do not send or receive 
\emph{any} data but get information about the network environment using ioctl().}}
\label{tab:udp_sockets_usage}
\end{table}

Looking at the amount of data sent/received over those UDP sockets, we
noticed that 85\% of the opened \texttt{SOCK\_DGRAM}
sockets do not send or receive \emph{any} data. Those sockets are
created to retrieve information about the networking environment using
\texttt{ioctl()}. While a single application opened 30\% of these
sockets, 15 applications use UDP and never send UDP data.
Table~\ref{fig:udp_ioctls} details the main \texttt{ioctl()}
requests. Although those \texttt{ioctl()} apply to any socket, we
suspect that applications perform them on UDP sockets because they are
cheaper than their TCP counterpart.

\begin{table}[]
    \centering
    \begin{tabular}{lll}
        \textbf{} & \textbf{Request} & \textbf{Purpose (get dev *)}      \\
        44\%   & SIOCGIFADDR      & Address                   \\
        25\%   & SIOCGIFNAME      & Name                      \\
        20\%   & SIOCGIFFLAGS     & Active flag word          \\
        5\%    & SIOCGIFNETMASK   & Network mask              \\
        5\%    & SIOCGIFBRDADDR   & Broadcast address         \\
        1\%    & Others           & N/A
    \end{tabular}
    \caption{ioctl() requests breakdown. 85\% of the UDP sockets use these
    requests to get information about the network devices.}
    \label{fig:udp_ioctls}
\end{table}

Overall, 16 applications sent or received data over UDP: 5 are
video-telephony apps such as Google Hangout or Skype, 4 are video
or music streaming applications such as Spotify or Netflix and 3 are
Google applications likely using QUIC like Chrome or Google Plus. The rest
are various applications that only exchange a few hundred bytes such as
Shazam or Angry Birds. Applications mainly use \texttt{sendto()} and
\texttt{recvfrom()} to send or receive data. We observed that 29\% of the
receiving calls set the \texttt{MSG\_PEEK} to peek on the receiving queue without
removing data and that 0.6\% of the sending calls set the \texttt{MSG\_NOSIGNAL}
flag to prevent a \texttt{SIGPIPE} from being raised in case of error.
We did not find any indication of the usage of the other flags on
\texttt{SOCK\_DGRAM} sockets. We noticed that Messenger uses the
\texttt{SIOCGSTAMP} iotcl during video calls roughly every second
\texttt{recvfrom()}. This ioctl allows round trip time measurements. Among the
\texttt{SOCK\_DGRAM} sockets that we observed, only 6\% sent or received data.
It is interesting to note that 8\% of the \texttt{SOCK\_DGRAM} sockets issued
a \texttt{connect()} without sending or receiving any data.

Multicast is one of the use cases for UDP sockets. We observed 8
applications that used UDP sockets to send multicast packets but
only 2 applications joined multicast groups. These 2 applications use
multicast to discover other similar applications on the same LAN, e.g.
using the Simple Service Discovery Protocol. A typical example are
streaming applications that allow to discover another device where
audio/video can be streamed over the network.

\section{TCP sockets}\label{sec:tcp}

Without much surprise, all our Android applications use TCP.
\texttt{SOCK\_STREAM} sockets account for 73\% of all opened
sockets. 63\% of these TCP sockets \texttt{connect()} to a remote
address while 37\% do not call \texttt{connect()} or \texttt{connect()}
to a loopback address. We first briefly analyze these later sockets
that interact with local daemons or applications. We then
analyze in more details the sockets that connect to distant servers.

\subsection{Local sockets}

\begin{figure}
\centering
\includegraphics[width=\columnwidth]{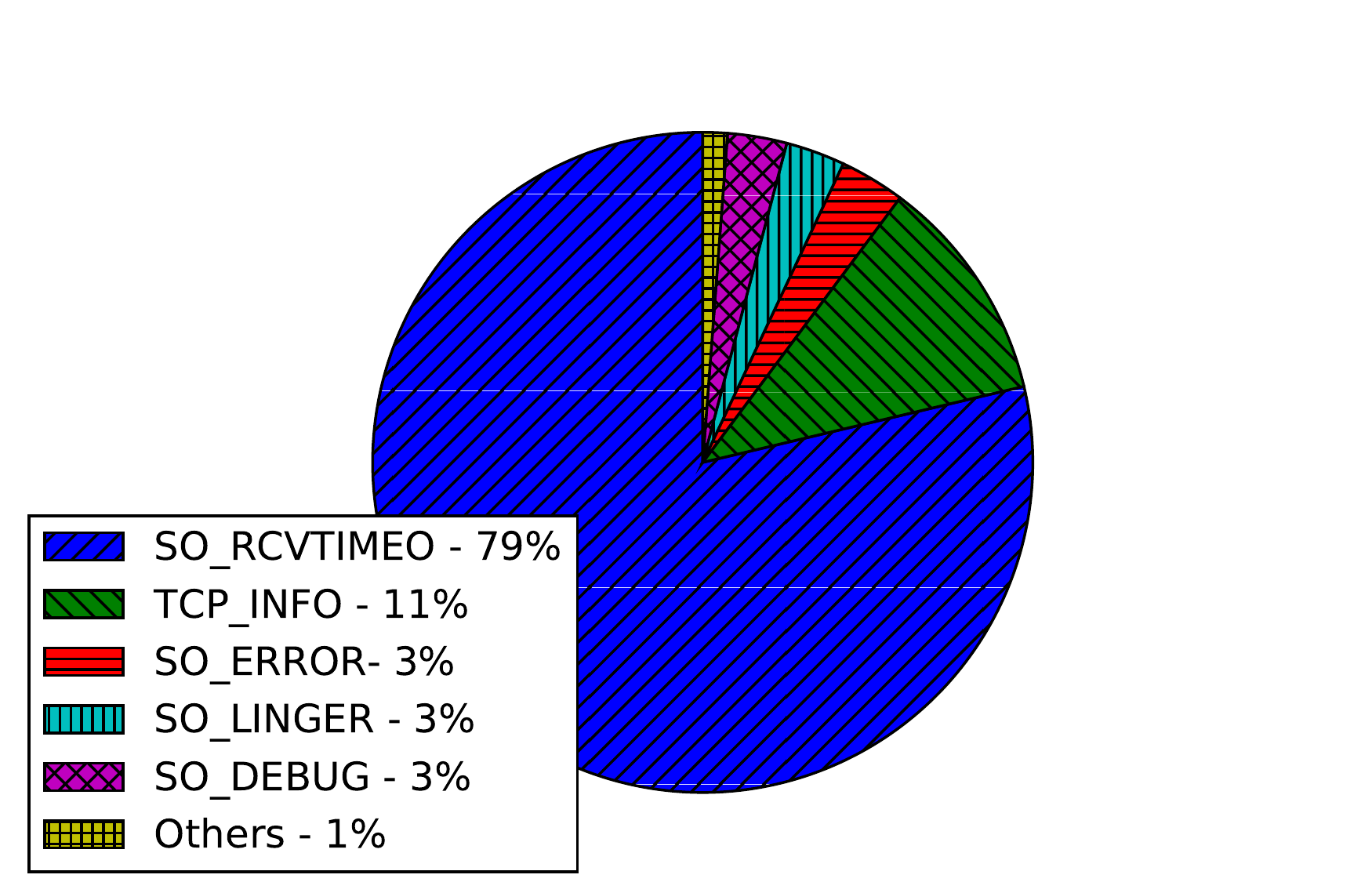}
\caption{getsockopt() and setsockopt() arguments for all TCP sockets (local
    and remote). SO\_RCVTIMEO is  by far the most used argument. SO\_ERROR is
    often used after a non-blocking connect(). The non-standard TCP\_INFO
    option is often retrieved. SO\_LINGER and SO\_DEBUG are often used together
    before a close() call.
}
\label{fig:sockopts_usage}
\end{figure}

We observe that a staggering 73\% of the local sockets only call
\texttt{setsockopt(SO\_RCVTIMEO)} once or several times after the initial
\texttt{socket()} call. As a result, figure~\ref{fig:sockopts_usage} shows how
the \texttt{SO\_RCVTIMEO} socket option dominates the \texttt{setsockopt()} and
\texttt{getsockopt()} arguments for TCP sockets (both local and remote). Since
\texttt{SO\_RCVTIMEO} only modifies the receiving timeout of the target socket,
these operations seem wasteful but we could not find a valid explanation for
this behavior. Another 16\% of the local sockets only call \texttt{close()}
after \texttt{socket()} and 3\% call \texttt{ioctl(SIOCGIWNAME)} to determine
if the current interface is wireless before closing the socket.
Table~\ref{tab:tcp_sockets_usage} summarizes these findings. While 85\% of the
UDP sockets use \texttt{ioctl()} to retrieve information about the network, we
rarely observe \texttt{ioctl()} on TCP sockets. This supports our
observation that applications prefer to perform \texttt{ioctl()} requests
on UDP sockets because they are less costly.

\begin{table}[]
\centering
\begin{tabular}{ll}
\hline
\textbf{37\%} & \textbf{Local sockets}  \\ \hline
27\%                                & \texttt{setsockopt(SO\_RCVTIMEO)}          \\
6\%                                 & Immediate \texttt{close()}                 \\
3\%                                 & Determine if interface is wireless         \\
1\%                                 & Other usages                               \\ \hline
\textbf{63\%} & \textbf{Remote sockets} \\ \hline
59\%                                & Exchange data after \texttt{connect()}     \\
4\%                                 & Do not send/recv data from network
\end{tabular}
\caption{TCP sockets usage. \textmd{37\% do not connect()
or connect() to a loopback address while 63\% connect() to a remote address.
Most local sockets only call setsockopt(SO\_RCVTIMEO).}}
\label{tab:tcp_sockets_usage}
\end{table}

\subsection{Remotely connected sockets}

\begin{figure}
\centering
\includegraphics[width=\columnwidth]{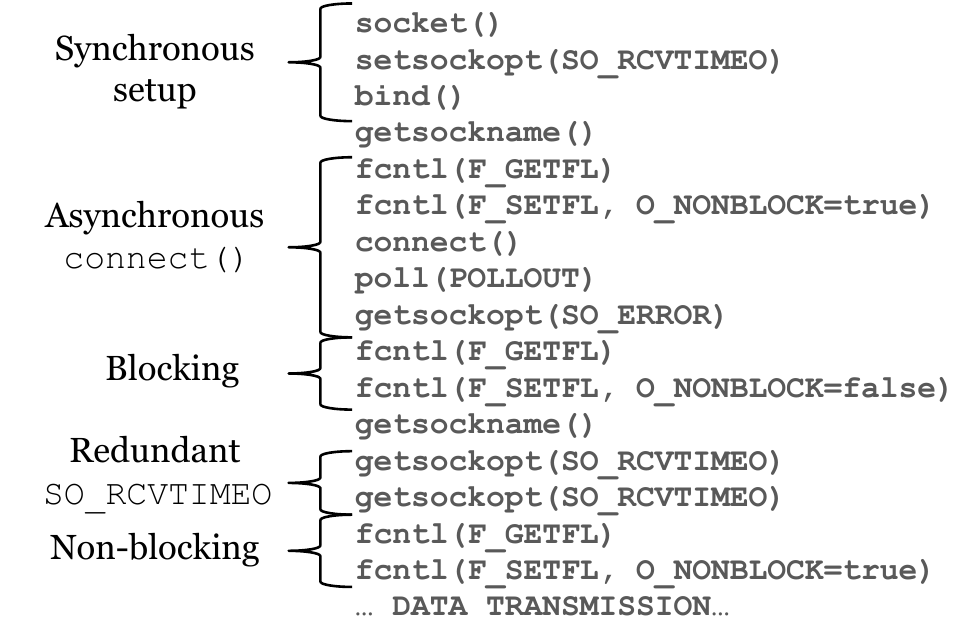}
\caption{Opening pattern on TCP sockets. \textmd{After a synchronous setup
phase that binds the socket, a non-blocking connect() call is issued. After 2
redundant getsockopt(SO\_RCVTIMEO), the socket is turned in non-blocking mode
again before the transmission of data.}}
\label{fig:opening_pattern}
\end{figure}

We now restrict our analysis to the 7505 TCP connections that
were used to contact a remote host. Various system calls could be used
to create those connections. However, our analysis reveals a common pattern of
16 socket API calls to open such a connection. This pattern is illustrated in
figure~\ref{fig:opening_pattern}. It results from the interactions
between the IO part of the Java Android core library \cite{aosp_socket}
and the \texttt{okhttp} external library \cite{aosp_okhttp_socketconnector}.
We first observe a synchronous setup phase that binds the socket.
The \texttt{set\-sock\-opt(SO\_RCV\-TIMEO)} call is issued by the
OkHTTP library. Then, the socket is put in non-blocking mode before
the \texttt{connect()} call. The successive \texttt{fcntl()} calls
modify the \texttt{O\_NONBLOCK} bit while keeping the values of the  other
flags. The \texttt{getsockopt (SO\_ERROR)} call checks whether the
\texttt{connect()} succeeded. Then the socket is turned synchronous again and
we observe two redundant calls to \texttt{getsockopt(SO\_RCVTIMEO)}, probably
related to the TLS library \cite{aosp_conscrypt}. Finally, the socket is put in
non-blocking mode again before the TLS handshake. The two \texttt{getsockname()}
calls are issued by the Android Java \texttt{Socket} to cache the local address
before and after the \texttt{connect()} call.

Surprisingly, 15 applications use the
\texttt{listen()} call. Among those, only 11 ever accepted an incoming
connection with \texttt{accept()} or \texttt{accept4()}. Among the 449 incoming
connection observed, 98\% originated from a loopback address.
We noticed 5 connections originating from a link-local address while
3 connections originated from a remote network. These 3 remote incoming
connections were accepted by a single application, Skype, that uses NAT
traversal.

\begin{figure}
\centering
\includegraphics[width=\columnwidth]{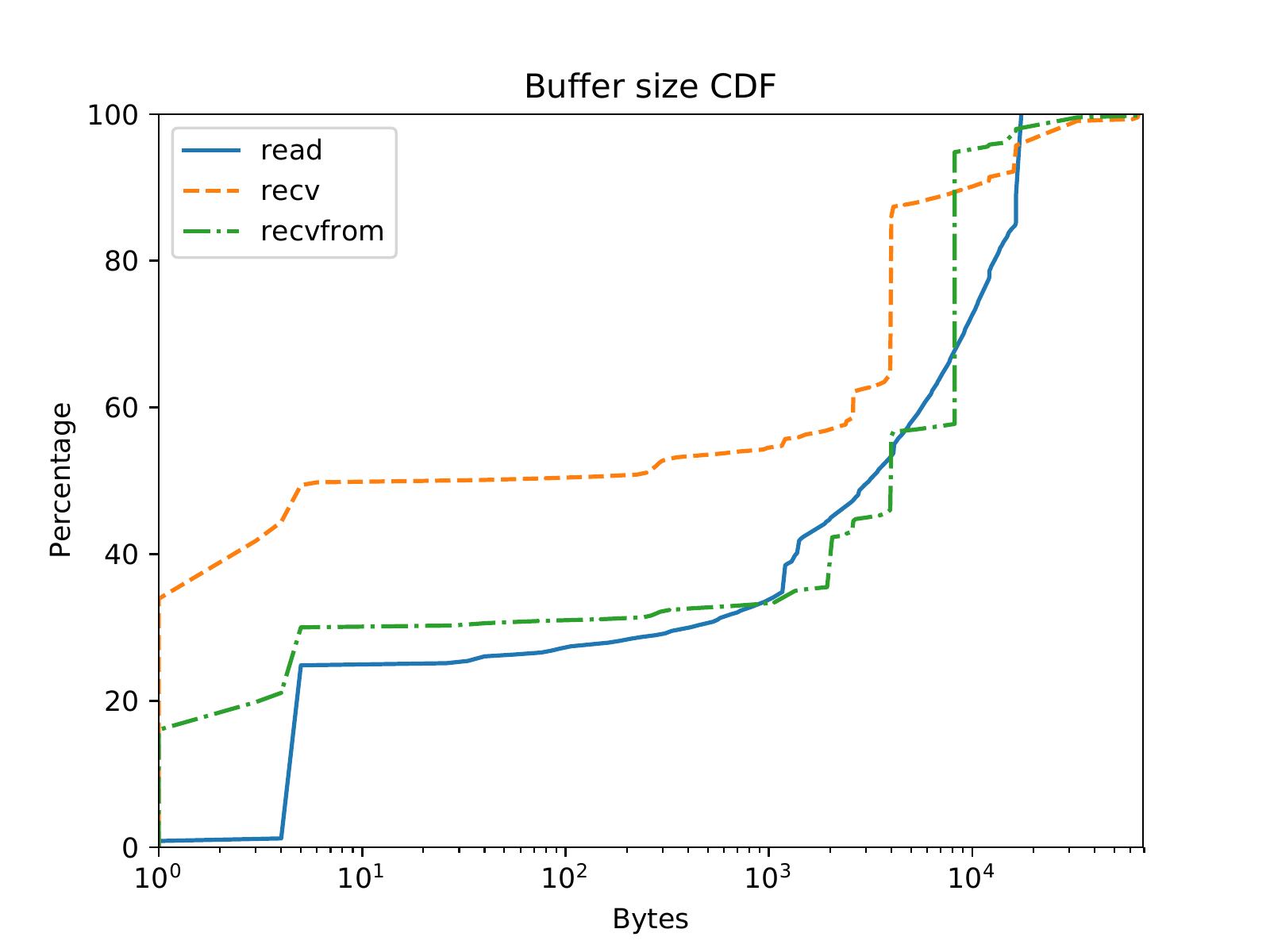}
\caption{Cumulative distribution function of the buffer size passed to the
        different receive functions. Half of the recv() calls use a buffer of 5
        bytes or less.}
\label{fig:recv_buffers}
\end{figure}

Let us now focus our analysis on the data transfer. 94\% of the TCP sockets
exchange data after the \texttt{connect()} call. Almost all
applications use the generic \texttt{read()} and \texttt{write()} calls.
Only half of them use their dedicated socket counterparts, \texttt{recv()} and
\texttt{send()}. Given the cost of issuing system calls, networking
textbooks recommend to use large buffers when transferring data.
\tcpsnitch allows to dissect how applications use each call.
Figure~\ref{fig:recv_buffers} shows a cumulative distribution function for the
size of the buffer given the the various receive functions. Surprisingly,
we observe that respectively 34\% and 16\% of the \texttt{recv()} and
\texttt{recvfrom()} calls use a buffer of exactly 1 byte and we also observe a
lot of 5 bytes long buffers. Overall, about half of the \texttt{recv()} calls
are passed a buffer of 5 bytes or less.

These functions support optional flags. The most popular sending flag is
\texttt{MSG\_NOSIGNAL} which is set on 60\% of the calls. This flag requests
not to send the \texttt{SIGPIPE} signal, which by default terminates the
process, when an application writes to a disconnected socket. It
is particularly useful for libraries since this flag does not modify
the process signal handlers. Only two other sending flags are used:
\texttt{MSG\_DONTWAIT} and \texttt{MSG\_MORE}. 13\% of the calls
are non-blocking thanks to the \texttt{MSG\_DONTWAIT} flag. 
\texttt{MSG\_MORE} is set on 2\% of the calls to indicate that more data is
coming. The other sending flags\footnote{\texttt{MSG\_CONFIRM},
\texttt{MSG\_DONTROUTE}, \texttt{MSG\_EOR} and \texttt{MSG\_OOB}} are never used.
18\% of the receiving calls are turned non-blocking using
the \texttt{MSG\_DONTWAIT} flag and 16\% of the calls set the
\texttt{MSG\_PEEK} flag to peek on the TCP receive queue without removing data.
Finally, a tiny fraction of those receiving calls (0.04\%) set
\texttt{MSG\_WAITALL} to request the operating system to block until it has
enough data to fill the buffer. The remaining flags\footnote{\texttt{MSG\_CMSG\_CLOEXEC},
\texttt{MSG\_ERRQUEUE}, \texttt{MSG\_OOB}, \texttt{MSG\_TRUNC}} do not
appear in our traces.

As observed for the connection establishment, there is a very frequent pattern
for the termination of a connection. 78 applications and about half of all
opened sockets use \texttt{getsockopt(SO\_DEBUG)} and
\texttt{getsockopt(SO\_LINGER)} before issuing \texttt{close()}. The
utilization of \texttt{SO\_DEBUG} at this point of the connection is
surprising. We investigated the Android source code and confirmed its usage in
the IO Java core library of Android \cite{aosp_blockguardos} where a function
closes all file descriptors. Because sockets using \texttt{SO\_LINGER} need
some additional processing to avoid the socket API \texttt{close()} call to
block, a \texttt{getsockopt()} is issued to detect if the file descriptor is a
socket. If this call succeeds, then the file descriptor is indeed a socket.
It seems that a failed \texttt{getsockopt(SO\_DEBUG)} is less critical from a
performance viewpoint than a failed \texttt{getsockopt(SO\_LINGER)},
hence its use. This closing pattern would certainly be observed
on a much higher proportion of the sockets if \tcpsnitch
could terminate cleanly the traced Android applications.

\begin{figure}
\centering
\includegraphics[width=\columnwidth]{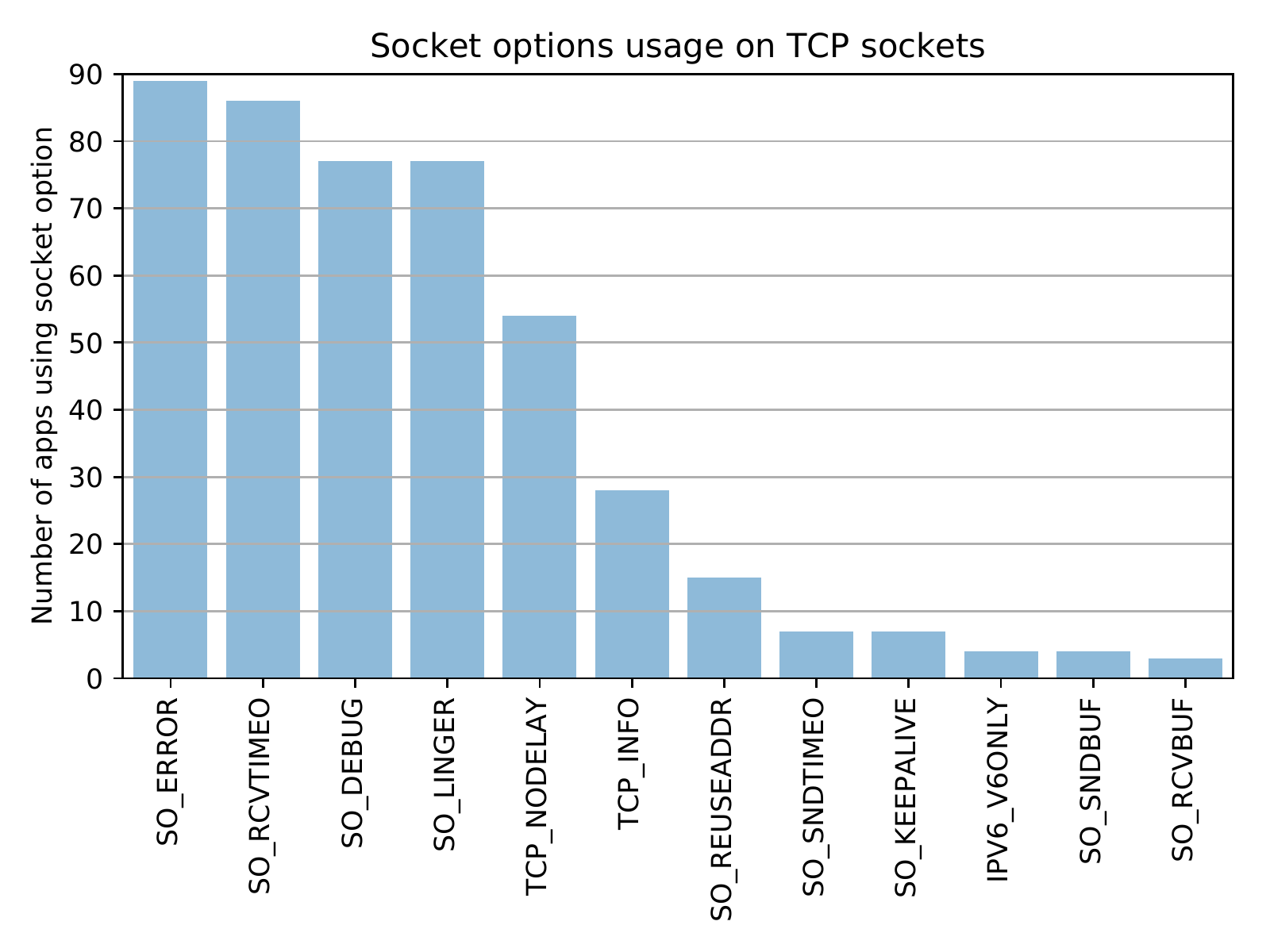}
\caption{Number of applications using each socket option.
    SO\_ERROR is often used after a non blocking
connect(). SO\_RCVTIMEO appears at the beginning of most TCP connections.
SO\_DEBUG and SO\_LINGER are used together before close(). TCP\_INFO
is used by a surprisingly large number of applications.}
\label{fig:sockopts_bars}
\end{figure}

\subsection{Socket options}

Socket options can be used by applications to tune the behavior of
the underlying TCP/IP stack. Linux supports a growing number of
non-standard socket options. Figure~\ref{fig:sockopts_bars} shows how many
applications use the main socket options observed in our dataset.
Several of these options are expected and some were discussed earlier,
\texttt{TCP\_INFO} was more surprising. This non-standard Linux TCP option
exports to the application counters maintained by the TCP stack. The standard
Android Java API does not expose this socket option and applications must
resort to a C/C++ library to use it. Still, 28 applications make use
of this socket option. As expected, those are mostly highly popular
applications such as Youtube, Chrome, Facebook or Spotify. For these
applications, \texttt{TCP\_INFO} was retrieved by 26\% of the
\texttt{SOCK\_STREAM} sockets and 73\% of these sockets retrieve
\texttt{TCP\_INFO} only once. Facebook, Messenger and Instagram are the
only applications that issue dozens of \texttt{TCP\_INFO} on a 
single TCP connection. For instance, we observed a Facebook TCP
connection lasting 32 seconds where \texttt{TCP\_INFO} was retrieved about
3000 times, almost as often as the 3500 \texttt{recv()} calls on the same
connection. These \texttt{TCP\_INFO} calls do not specifically happen at the
start or the end of a connection, but seem uniformly distributed during the
lifetime of the TCP connection. As figure~\ref{fig:sockopts_usage} shows,
\texttt{TCP\_INFO} is the second most used socket option argument for TCP
sockets.

\section{Discussion}\label{sec:conclusion}

We have proposed \tcpsnitchns, an application that intercepts network
system and library calls on the Linux and Android platforms to
collect more information about their usage, including the parameters
passed to those API calls. We collected more than 2.3 millions calls made by
90 popular applications on sixteen thousands sockets. The application and the
collected dataset are publicly available\footnote{The entire dataset can be
explored via \url{https://tcpsnitch.org}. The \tcpsnitch source code is
available from \url{https://github.com/GregoryVds/tcpsnitch} and the web
interface can be retrieved from \url{https://github.com/GregoryVds/tcpsnitch_web}.
}.

Our analysis revealed several interesting patterns for the utilization
of the socket API on Android applications. First, in an IPv6 enabled
WiFi network, these applications prefer IPv6 over IPv4. Second, UDP sockets are
mainly used as a shortcut to retrieve information about the network
configuration. Third, many Android applications use the same pattern
of system calls to establish and terminate TCP connections. Fourth, Android
applications use various socket options, even some like
\texttt{TCP\_INFO} that are not directly exposed by the standard Java
API.

\tcpsnitch and its associated website already provide a good overview
of how real applications use the socket API. Our future work will be
to add traces from more applications in the database and support other
platforms starting with MacOS.

\newpage
\bibliographystyle{plain}
\bibliography{tcpsnitch}

\end{document}